\documentclass[aps,prd,twocolumn,superscriptaddress,nofootinbib]{revtex4-2}

\usepackage[T1]{fontenc}
\usepackage[utf8]{inputenc}
\usepackage[american]{babel}
\usepackage{epsfig}
\usepackage{xspace}
\usepackage{graphicx}
\usepackage{booktabs}
\usepackage{multirow}
\usepackage{dcolumn}
\usepackage{amsmath}
\usepackage[dvipsnames]{xcolor}
\usepackage{subfigure}
\usepackage{mathtools}
\usepackage{amsfonts}
\usepackage{amssymb}
\usepackage{epstopdf}
\usepackage{siunitx}
\usepackage{braket}
\usepackage{enumitem}
\usepackage{soul}
\usepackage{hyperref}
\hypersetup{
	colorlinks=true, % false: boxed links; true: colored links
	linkcolor=blue, % color of internal links (change box color with linkbordercolor)
	citecolor=Red}

\begin{document}

\title{Wave-particle duality and entanglement in neutrino oscillation }

\author{Rajrupa Benerjee}
\email{rajrupab@iitbhilai.ac.in}
\affiliation{Department of Physics, Indian Institute of Technology Bhilai, Durg 491001, India}

\author{Pratidhwani Swain}
\email{ps.rs.phy@buodisha.edu.in}
\affiliation{Department of Physics, Berhampur University, Bhanja Bihar, Berhampur, Odisha,760007, India}
\affiliation{Department of Physics, Nimapara Autonomous College, Nimapara, Odisha, 752106, India}

\author{Prasanta K. Panigrahi}
\email{panigrahi.iiser@gmail.com}
\affiliation{Center for Quantum Science and Technology, Siksha ’O’ Anusandhan, Bhubaneswar-751030, India}
\affiliation{Indian Institute of Science Education and Research Kolkata, 741246, WB, India}

\author{Sudhanwa Patra}
\email{sudhanwa@iitbhilai.ac.in}
\affiliation{Department of Physics, Indian Institute of Technology Bhilai, Durg 491001, India}
\affiliation{Institute of Physics, Bhubaneswar, Sachivalaya Marg, Bhubaneswar 751005, India}

%\date{\today}

\begin{abstract}
%\textcolor{red}{We investigate neutrino oscillations from the perspective of a three-slit interference analogue, which naturally leads to the recently established triality condition linking wave–particle duality and entanglement. Within this framework, the optical counterparts of visibility, path predictability, and entanglement are explicitly constructed, and their dynamical evolution in neutrino oscillations is analyzed. For both two- and three-flavor systems, we demonstrate that the evolution yields distinct energies at which entanglement vanishes, as well as points where maximal entanglement emerges. The role of entanglement monogamy in constraining flavor correlations is highlighted, together with its implications for neutrino measurements and quantum information aspects of oscillation phenomena.}
We investigate wave--particle--entanglement complementarity in
three-flavor neutrino oscillations within a quantum
information--theoretic framework. Treating neutrino flavor evolution as
an open quantum system and explicitly accounting for
detector--propagation correlations, we extend the conventional
wave--particle duality relation to a triality relation involving
predictability, visibility, and entanglement. Using reduced density
matrices and I-concurrence as a quantitative measure of entanglement, we
demonstrate that the total information content of the system satisfies
the relation $\mathcal{P}^2 + \mathcal{V}^2 + \mathcal{E}^2 = 1$. While
predictability and visibility exhibit the expected complementary
behavior, we show that entanglement encodes additional wave-like
information that is not captured by visibility alone. We apply our
formalism to realistic long-baseline neutrino experiments, namely \textsf{DUNE}
and \textsf{T2K}, and find that at the first oscillation maximum, a simultaneous
characterization of the particle-like and wave-like nature of neutrinos
becomes possible through the combined measurement of predictability and
entanglement. Our results provide a unified operational interpretation
of neutrino oscillations and highlight the role of quantum correlations
in extending wave--particle duality to multipartite systems.
\iffalse
We analyze neutrino oscillations from the perspective of quantum complementarity by extending the familiar wave–particle duality into an exact triality relation among visibility, predictability, and entanglement. Using a density-matrix approach, we construct explicit measures of these quantities in both two and three-flavor frameworks and show that they satisfy an exact conservation-like identity throughout oscillation evolution. In this picture, visibility corresponds to interference strength, predictability to flavor imbalance, and entanglement to the reduction of purity due to flavor correlations. 
We demonstrate that neutrino oscillations naturally exhibit regimes of maximal entanglement, vanishing coherence, and entanglement monogamy constraining flavor correlations. Our results establish that long-baseline neutrino oscillations realize the triality relation in an energy-dependent way, with \textsf{T2K} probing vacuum-like coherence-entanglement balance and \textsf{DUNE} revealing the decisive role of matter effects in redistributing the quantum information budget.
\fi
\end{abstract}

\maketitle

\section{Introduction}
\label{sec1}
 The seminal disagreement between Einstein and Bohr over the nature of quantum reality has revolved around the limits of wave-particle duality and Bohr's principle of complementarity. Einstein argued that a sufficiently refined measurement, such as his recoiling-slit thought experiment, could, in principle, reveal both particle-like trajectories and wave-like interference, challenging the completeness of quantum mechanics \cite{Zhang:2024rts}. Bohr countered that any attempt to acquire path information inevitably destroys interference, a view now strongly supported by modern high-precision experiments. 
 Since the early days of quantum theory, the concept of complementarity, introduced by Niels Bohr in 1928, has shaped our understanding of quantum behavior \cite{Bohr:1935af}.
 
The idea that wave-like and particle-like properties cannot be simultaneously revealed soon evolved into quantitative formulations that provided one of the first rigorous analyses \cite{Wootters:1979zz}. This was followed by a study that introduced a connecting path between the particle nature and wave nature in terms of predictability $\mathcal{P}^2$ and visibility $\mathcal{V}^2$ \cite{greenberger1988}. 
\begin{eqnarray}
    \mathcal{P}^2 + \mathcal{V}^2=1
\end{eqnarray}
Extensive literature has refined this duality relation, demonstrating that the more knowledge one has about the particle's path, the weaker its interference pattern becomes.

However, the familiar duality relation reaches equality only for perfectly pure and isolated systems. In realistic scenarios, especially when the system becomes correlated or entangled with other degrees of freedom, the equality no longer holds, i.e., $\mathcal{P}^2 + \mathcal{V}^2 \leq 1$ \cite{jaeger1995two, englert1996}. This observation demonstrates that predictability and visibility alone do not account for all the quantum information present in the system. To fully balance the relation, a third contribution is required, one that captures the correlations or entanglement shared with other degrees of freedom \cite{Jakob:2007zz}. Introducing this additional term elevates the duality relation to a triality relation, providing a more complete and consistent description of quantum complementarity.
The next significant step came with attempts to extend complementarity beyond two-path systems. However, soon it became clear that predictability and visibility alone cannot account for all quantum features in multi-path scenarios~\cite{Bera:2015wbe, qureshi2017wave, Bagan:2017dtn, PhysRevLett.86.559}. The missing element is entanglement, which fundamentally ties together $\mathcal{P}^2$ and $\mathcal{V}^2$. 
 Within this conceptual backdrop, quantum triality emerges as a natural extension of the traditional dualistic viewpoint. Rather than treating wave-like, particle-like, and information-theoretic aspects as mutually exclusive descriptions, triality emphasizes their co-emergence within a unified quantum structure. It generalizes complementarity by recognizing that quantum systems simultaneously encode coherence (wave aspect), discreteness (particle aspect), and correlation or entanglement (informational aspect), each of which becomes manifest depending on the measurement context. This perspective situates triality as a deeper structural principle that captures the informational, dynamical, and relational facets of quantum phenomena.
The familiar duality relation may be generalized to a triality framework in which predictability, visibility, and I-concurrence (or an equivalent entanglement measure) jointly characterize the system in the following form~\cite{Hill:1997pfa,Roy:2021wcg},
%%%%%%%%
\begin{eqnarray}
    \mathcal{P}^2+\mathcal{V}^2+\mathcal{E}^2=1
\end{eqnarray}
which demonstrates that entanglement $\mathcal{E}^2$ is the essential quantity completing the balance among the complementary aspects of quantum behavior \cite{Basso:2022ihl}.
Subsequent work extended this idea to multi-path and multipartite systems, emphasizing that complementarity is not merely a competition between particle-like and wave-like tendencies but also involves the correlations a system shares with its environment or ancillary degrees of freedom \cite{Qureshi:2021kgy}.

On the other hand, as neutrinos travel, their flavor composition changes through coherent superpositions of mass eigenstates, leading to the familiar appearance and disappearance probabilities measured in experiments ~\cite{Bilenky:2004xm, Kayser:2012ce, SNO:2002tuh, T2K:2011ypd, MINOS:2011amj, K2K:2006yov, DayaBay:2012fng, Super-Kamiokande:2002ujc, NOvA:2016vij, DUNE:2020jqi}. While the global neutrino state remains pure during this evolution, the individual flavor components behave like subsystems that can lose purity once a specific detection flavor is selected. In practical terms, tracing out any flavor degree of freedom yields a reduced state with $\mathrm{Tr}[\rho_{_{\beta\gamma}}^{\alpha^2}] < 1$, signaling that the subsystem has become mixed ~\cite{Bittencourt:2023asd, Jakob:2007zz}.
This intrinsic interplay between purity, coherence, and mixedness makes neutrinos a natural platform for applying quantum information concepts. Ideas such as coherence, entanglement, and complementarity-traditionally examined in optical or atomic systems-find a meaningful and experimentally accessible realization in the context of three-flavor neutrino oscillations~\cite{Baumgratz:2013ecx, Bittencourt:2023asd, Roy:2021wcg}. The quantum foundations of the neutrino system within the framework of neutrino oscillation have been explored from several complementary perspectives. Earlier studies examined the entanglement among flavor and mass modes~\cite{Blasone:2007wp, Blasone:2010ta, Banerjee:2015mha}, while others investigated the non-classical nature of oscillations using Bell inequalities and Leggett-Garg inequalities~\cite{Gangopadhyay:2013aha, Gangopadhyay:2017nsn, Naikoo:2019eec}. Additionally, some efforts are made on quantum coherence ~\cite{Bittencourt:2023asd}, including steering~\cite{Konwar:2024nrd} and entropic uncertainty relations~\cite{Blasone:2022joq, Blasone:2010ta, Jha:2020dav, Cirigliano:2024pnm}, to better understand the quantum resources present during neutrino propagation. Together, these investigations highlight the rich quantum structure that underlies oscillation phenomena and motivate the use of quantum-information tools in their analysis~\cite{konwar2025effectoffdiagonalnsiparameters, Jha:2022yik, Jha:2025ekn, jha2025quantumspreadcomplexityprobe, Bouri:2024kcl, Siwach:2022xhx, Siwach:2024jet}.

From this perspective, neutrino oscillations offer a particularly appealing arena to explore the broader idea of complementarity \cite{Bittencourt:2023asd}. 
Verification of quantum triality has recently been done in the photon system. 
Verification of the quantum triality relation in neutrino oscillations has been explored in earlier work, mainly through studies of internal flavor-mode entanglement. In these analyses, visibility often vanishes, effectively reducing the triality relation to the usual duality relation \cite{Bittencourt:2023asd}. However, a comprehensive treatment that incorporates entanglement between the detected flavor state and the remaining propagation modes has not yet been systematically carried out. Questions such as when the triality relation genuinely holds for neutrino oscillations and under what circumstances it collapses back into duality remain largely open and have received limited attention. These gaps motivate a more complete investigation of detector-state entanglement within the three-flavor framework. 

In this context, the triality relation offers a unified framework for characterizing the accessible information in three-flavor oscillations. Predictability quantifies the particle-like distinguishability of flavor outcomes; visibility captures the wave-like interference among propagation amplitudes \cite{Roy:2021wcg, Rungta:2001zcj}; and entanglement, defined here through a generalized I-concurrence \cite{Bhaskara:2017uzs}, measures the nonlocal correlations between the detected flavor and the remaining flavor modes. Together, these quantities provide a complete description of how information about neutrino flavor is distributed during propagation.

In this work, we develop a theoretical framework for the quantum triality relation in neutrino oscillations, linking predictability $\mathcal{P}^2$, visibility $\mathcal{V}^2$, and entanglement $\mathcal{E}^2$ through a generalized I-concurrence. We show that when the entanglement between the detected flavor and the propagating flavor modes is correctly accounted for, the triality condition is fully satisfied in the three-flavor neutrino system. In contrast, analyses that consider only internal flavor-mode entanglement are unable to recover the complete triality structure. A comparison with the two-flavor framework further clarifies this distinction. Tracing out the detector state leaves a single remaining degree of freedom, reducing the system to a single-partite configuration where the complementarity relation naturally collapses to the familiar duality relation.

Building on this formalism, we investigate the energy-dependent behavior of the triality components in realistic experimental scenarios. Using simulations performed with \textsf{GLoBES} (General Long Baseline Experiment Simulator)~\cite{Huber:2004ka, Huber:2007ji}, we study long-baseline accelerator experiments such as \textsf{DUNE} (Deep Underground Neutrino Experiment)~\cite{DUNE:2020jqi} and \textsf{T2K} (Tokai to Kamioka)~\cite{T2K:2011ypd}, whose different baselines and energy spectra allow us to examine triality in both vacuum-dominated and matter-influenced regimes. 
We demonstrate that the triality relation is realized within the experimental setups of both experiments.
Our analysis reveals that the triality relation remains robust across these settings, capturing the essential quantum information structure of three-flavor oscillations. This framework not only deepens the conceptual understanding of flavor evolution but also creates opportunities for connecting neutrino physics with broader themes in quantum information and quantum foundations.

The structure of this paper is as follows. In Sec.~\ref{neuosc}, we present a brief overview of three-flavor neutrino oscillations using the density-matrix formalism. Section~\ref{trialityneu} introduces the triality framework and develops the expressions for $\mathcal{P}^2$, $\mathcal{V}^2$, and $\mathcal{E}^2$ within this formalism. Building on this, Sec.~\ref{experiment} describes the experimental analysis, where we test the triality relation using \textsf{GLoBES} simulations for two long-baseline experiments, \textsf{DUNE} and \textsf{T2K}. We close the paper in Sec .~\ref {conclusion} with a summary of our findings and their broader implications.
\section{Neutrino Oscillations in the Density Matrix Framework}
\label{neuosc}
The coherent mixing of flavor and mass eigenstates governs neutrino oscillations. The flavor eigenstates
$\ket{\nu_\alpha}$ $(\alpha=e,\mu,\tau)$ produced at the source can be written as a superposition of mass eigenstates  $\ket{\nu_j}$ $(j=1,2,3)$ as
\begin{equation}
 \ket{\nu_\alpha}  
\;=\; \ket{\nu_{\alpha}(L=0)}
\;=\; \sum_{j=1}^3 U^*_{\alpha j} \ket{\nu_j}\;.
\end{equation}
where $U$ is the Pontecorvo-Maki-Nakagawa-Sakata (PMNS)~\cite{Pontecorvo:1967fh, Patra:2023ltl} mixing matrix parametrised in terms of neutrino mixing angles ($\theta_{ij}$) and Dirac CP pahse ($\delta$) as
\begin{eqnarray}
U 
& = & R_{23}(\theta_{23},0)\;R_{13}(\theta_{13},\delta)\;R_{12}(\theta_{12},0)
\phantom{\bigg|} \cr
&&\hspace{-1cm} =
\left[ \begin{array}{ccc}
c_{12}c_{13} & s_{12}c_{13} & s_{13} e^{-i\delta} \\
-s_{12}c_{23} - c_{12}s_{13}s_{23}e^{i\delta} &
\phantom{-}c_{12}c_{23} - s_{12}s_{13}s_{23}e^{i\delta} & c_{13}s_{23} \\
\phantom{-}s_{12}s_{23} - c_{12}s_{13}c_{23}e^{i\delta} &
-c_{12}s_{23} - s_{12}s_{13}c_{23}e^{i\delta} & c_{13}c_{23}
\end{array} \right] \nonumber
\label{UPparam}
\end{eqnarray}
Here, $R_{ij}(\theta,\delta)$ denotes a rotation matrix and 
$s_{ij}\equiv\sin\theta_{ij}$, $c_{ij}\equiv\cos\theta_{ij}$.
Neutrino oscillations can be significantly altered in the presence of neutrino interaction with matter. This modification arises due to the coherent forward scattering of neutrinos with particles in the medium, which introduces a potential term, $V_{CC}$, given by $V_{CC}=\sqrt{2}G_{F}N_{e}$ \cite{PhysRevD.17.2369, MIKHEYEV198941, Bethe:1986ej, Smirnov:2004zv}, where $G_{F}$ is the Fermi coupling constant and  $N_{e}$ represents the electron number density.
The effective Hamiltonian governs the propagation in the flavor basis, read as,
\begin{equation}
   \mathcal{H}=\frac{\Delta m_{31}^{2}}{2E}\big[U \mbox{diag}(0,\alpha,1)U^{\dagger}+\mbox{diag}(\Hat{A},0,0)\big],
\label{eq:2.1}
\end{equation}
where $\alpha=\Delta m_{21}^{2}/\Delta m_{31}^{2}$, $\Hat{A}=A/\Delta m^{2}_{31}$ 
with $A=2 V_{CC}$ is representing matter-induced interactions. Here, $\Delta m_{21}^2$ and $\Delta m_{31}^{2}$ are the solar and atmospheric mass square differences, respectively, and $E$ is the energy of the neutrino.
For example, this leads to the construction of the time-evolved flavor state taking $\nu_{\mu}$ as the initial flavor state,
\begin{equation}
    \ket{\nu_{\mu}(t)}=\Tilde{U}_{\mu e}\ket{\nu_e}+\Tilde{U}_{\mu\mu}\ket{\nu_\mu}+\Tilde{U}_{\mu \tau}\ket{\nu_\tau},
\label{eq:state}
\end{equation}
where, $\Tilde{U}_{\alpha\beta}=U_{\beta i}U_{\alpha i}^{*}e^{-i E_{i}t}$, $E_i=m^{2}_{i}/2E$ (i=1,2,3) is the eigenvalue of the mass basis Hamiltonian. Within the ultrarelativistic approximation, the propagation time can be
identified with the baseline length, $ t\simeq L$. In this limit, the dimensionless combination $\Delta m^{2}_{ij} L / 4E$ determines the characteristic oscillation frequency governing neutrino flavor transitions.
%The time-evolved state is obtained from the unitary operator $U(t) = e^{-i\,H\,t}$.
The density matrix formalism provides a convenient description of oscillations in the context of quantum complementarity. In the three-generation neutrino oscillation framework, we can reinterpret the flavor states in terms of occupation numbers, $ \ket{\nu_{e}} \equiv \ket{1}_{e}\otimes\ket{0}_{\mu}\otimes\ket{0}_{\tau}$, $ \ket{\nu_{\mu}} \equiv \ket{0}_{e}\otimes\ket{1}_{\mu}\otimes\ket{0}_{\tau}$ and $ \ket{\nu_{\tau}} \equiv \ket{0}_{e}\otimes\ket{0}_{\mu}\otimes\ket{1}_{\tau}$ \cite{Konwar:2024nrd,Li:2021epj,Blasone:2014cub,KumarJha:2020pke}. This perspective motivates a further investigation into neutrino entanglement, particularly in the context of the neutrino experiments.  In this scenario, the density matrix for the state $\ket{\nu_{\mu}(t)}$, defined as 
$$\rho^{\mu}_{e\mu\tau}(t)=\ket{\nu_{\mu}(t)}\bra{\nu_{\mu}(t)}$$
which evolves unitarily as $\rho^{\mu}_{e\mu\tau}(t) = U(t)~\rho^{\mu}_{e\mu\tau}(0)~U^\dagger(0)$
where $\rho^{\mu}_{e\mu\tau}(0) = \ket{\nu_{\mu}}\bra{\nu_{\mu}} $
represents the initially produced flavor. The survival and transition probabilities are then the diagonal elements of the density matrix,
$P(t)_{\nu_\mu\rightarrow\nu_\alpha} = 
 \bra{\nu_{\alpha}} \rho_{e\mu\tau}^{\mu}(t) \ket{\nu_{\alpha}}~, \mathrm{where} \quad \alpha\in \left\{e,\mu,\tau\right\}~.$
This satisfies probability conservation identity $\sum_{\alpha} P_{\mu \to \alpha} =1$.

Let us assume that the initial flavor is muon neutrino $\nu_\mu$ and considering three flavor neutrino mixing analysis, the derived expression for dynamically evolved density matrix considering neutrino as a single qutrit system, is given by \cite{Konwar:2024nrd, Li:2021epj},  
\begin{eqnarray}
\rho^{\mu}_{e\mu\tau}(t)  =\begin{pmatrix}
          0 & 0 & 0 & 0 & 0 & 0 & 0 & 0\\
          0 & ~|\widetilde{U}_{\mu\tau}|^{2} ~ & ~\widetilde{U}_{\mu\tau}\widetilde{U}^{*}_{\mu\mu} ~&~ 0 ~&~ \widetilde{U}_{\mu\tau}\widetilde{U}^{*}_{\mu e} ~&~ 0 ~&~ 0 ~& 0\\
          0 &~ \widetilde{U}_{\mu\mu }\widetilde{U}^{*}_{\mu\tau} ~&~ |\widetilde{U}_{\mu\mu}|^{2} ~&~ 0 ~&~ \widetilde{U}_{\mu\mu}\widetilde{U}^{*}_{\mu e} ~&~ 0 & 0 & 0\\
          0 & 0 & 0 & 0 & 0 & 0 & 0 & 0\\
          0 & \widetilde{U}_{\mu e}\widetilde{U}^{*}_{\mu\tau} & \widetilde{U}_{\mu e}\widetilde{U}^{*}_{\mu\mu} & 0 & |\widetilde{U}_{\mu e}|^{2} & 0 & 0 & 0\\
          0 & 0 & 0 & 0 & 0 & 0 & 0 & 0\\
          0 & 0 & 0 & 0 & 0 & 0 & 0 & 0\\
          0 & 0 & 0 & 0 & 0 & 0 & 0 & 0\\
      \end{pmatrix}~.
      \nonumber\\
\label{eq:DM}
\end{eqnarray}
The elements of the time-evolved density matrix directly encode the appearance and disappearance probabilities of neutrinos, namely
\begin{equation}
|\widetilde{U}_{\mu\mu}|^{2}=P_{\mu\mu}, \qquad 
|\widetilde{U}_{\mu e}|^{2}=P_{\mu e}, \qquad 
|\widetilde{U}_{\mu\tau}|^{2}=P_{\mu\tau},
\end{equation}
which satisfy the normalization condition
\begin{equation}
|\widetilde{U}_{\mu\mu}|^{2}+|\widetilde{U}_{\mu e}|^{2}+|\widetilde{U}_{\mu\tau}|^{2}=1.
\end{equation}
\noindent
Although the full time-evolved neutrino state remains pure, 
\begin{equation}
\mathrm{Tr}\,\rho^\mu_{e\mu\tau}(t)=\mathrm{Tr}\,\left[\rho_{e\mu\tau}^\mu(t)\right]^2=1~,
\end{equation}
Its flavor components do not evolve independently of one another. Instead, the oscillation process generates quantum correlations among the three flavor modes, which can lead to nonzero entanglement within the flavor triplet. Thus, the purity of the global state does not restrict the presence of entanglement between individual flavor modes.

To diagnose this entanglement, we examine the reduced density matrix obtained by tracing out any one flavor mode. For neutrinos, such a reduced state always satisfies \cite{Bittencourt:2023asd}
\begin{equation}    \mathrm{Tr}\left[\rho_{\small{\alpha\beta}}^{\mu}\right]^2 < 1, \quad \mathrm{where} \quad \alpha,\beta\in\{e,\mu,\tau\}
\label{eq:reduce}
\end{equation}
indicating that the subsystem is mixed even though the whole state is pure. 
In Eq.~(\ref{eq:reduce}), the superscript $\mu$ denotes the initial flavor mode $\mu$ and $\{\alpha,\beta\}$ represents the remaining two propagating flavor modes apart from the traced out flavor mode. For 
this loss of purity reflects the fact that tracing out a correlated partner removes interference information, leaving behind classical statistical uncertainty. The mixedness of the reduced state, therefore, serves as a direct indicator of entanglement among the flavor modes.

While the global muon-neutrino state $\nu_\mu(t)$ remains pure and continues to satisfy the duality relation, introducing a detector sensitive to a particular flavor modifies the situation. Specifying a detector state, for example, $\nu_\mu$, effectively traces over that flavor and incorporates the contribution of neutrino---detector entanglement into the duality inequality. 
%In this sense, tracing out the $\nu_\mu$ component quantifies the entanglement shared between $\nu_\mu$ and the remaining $\{\nu_e,\nu_\tau\}$ subsystem.

Since long-baseline accelerator experiments predominantly measure $\nu_\mu$ disappearance and $\nu_e$ appearance, it is natural to choose $\nu_\mu$ as the detector-defined flavor mode for our analysis. We emphasize that two detector states cannot be considered simultaneously, because tracing out two flavor modes leaves a single-mode state, for which the three-flavor triality relation does not apply.

In the following subsection, we explicitly compute the predictability, visibility, and entanglement measures for the three-flavor neutrino system using $\nu_\mu$ as the detector state.
\section{Triality in Neutrino Oscillations}
\label{trialityneu}
Building on the discussion above, where reduced flavor states naturally acquire mixedness due to entanglement with the detector mode, we now introduce the framework of complementary relations, which provides a unified way to quantify the interplay between flavor information and flavor correlations in neutrino oscillations. Complementarity, originally formulated in the context of wave--particle duality, asserts that specific properties of a quantum system---though equally real---cannot be simultaneously accessed with arbitrary precision.

In the exemplary two-slit experiment, this duality is characterized by two quantities---the predictability $(\mathcal{P})$, which encodes the knowledge of path information of the particle, and the visibility $(\mathcal{V})$, which measures the sharpness of the interference pattern. These quantities satisfy the standard bound
\begin{equation}
    \mathcal{P}^{2} + \mathcal{V}^{2} \leq 1,
\end{equation}
with equality holding only for a pure single-partite state.
However, as established in the previous section, the flavor evolution of a neutrino, once a specific detector flavor is identified, effectively becomes a bipartite scenario. The global propagation state remains pure, but the reduced state associated with the detector flavor becomes mixed due to entanglement with the remaining flavor modes. In such cases, the usual wave-particle duality is incomplete as it omits the contribution of the entanglement effect introduced by tracing out the detector state.

To obtain a complete description of the system, the standard wave–particle
duality relation must be extended by introducing a third quantity,
$\mathcal{E}^2$, which quantifies the quantum entanglement between the
chosen detector flavor and the remaining flavor modes. This entanglement
also carries intrinsically wave-like information associated with the
neutrino system. The inclusion of $\mathcal{E}^2$ naturally leads to the
triality relation
\begin{equation}
\mathcal{P}^{2} + \mathcal{V}^{2} + \mathcal{E}^{2} = 1.
\end{equation}
Thus, beyond the conventional visibility $\mathcal{V}^2$, the wave nature
of neutrinos is additionally encoded in the entanglement measure
$\mathcal{E}^2$.
%This holds exactly for pure global states. 
Here, $\mathcal{P}^2$ and $\mathcal{V}^2$ represent complementary local aspects of the detected flavor mode, while $\mathcal{E}^2$ quantifies the information shared with the detector and an individual propagation mode. Such correlations naturally arise in neutrino oscillations, where the propagation dynamics continuously redistribute quantum correlations among the flavor modes.

In the remainder of this section, we show explicitly how the three-flavor neutrino system satisfies this quantum triality relation when $\nu_\mu$ is chosen as the detector state, thereby establishing a direct correspondence between oscillation probabilities, wave--particle duality, and flavor-mode entanglement.
\begin{itemize}
\item \pmb{Predictability ($\mathcal{P}^2$) :} Predictability, $\mathcal{P}^2$, quantifies how strongly a particular flavor outcome dominates over the others. In other words, it measures the degree of flavor imbalance, with larger values indicating a stronger preference for one flavor over the other. For an $n$-level system, predictability is defined in terms of the diagonal elements of the density matrix $\rho$ as \cite{PhysRevA.70.062107, PhysRevA.64.042113, Bittencourt:2023asd}
\begin{equation}
    \mathcal{P}^2 = \frac{n}{n-1}\sum_{i}\rho_{ii}^{2} - \frac{1}{n-1},
    \label{eq:predictability}
\end{equation}
which vanishes for a completely democratic state (where all outcomes are equally likely) and approaches a maximum (unity) when a single outcome dominates.
Where $\rho_{ii}$ is the diagonal element of the density matrix.
This definition integrates naturally into the triality framework introduced above. From a physical standpoint, the predictability $\mathcal{P}^2$ quantifies the particle-like information encoded in the detected flavor mode.
In the case of a three-flavor neutrino system, however, Eq.~(\ref{eq:DM}) cannot be directly employed to evaluate $\mathcal{P}^2$, since the flavor density matrix $\rho^{\mu}_{e\mu\tau}(t)$ corresponds to a pure state and therefore does not capture the relevant triality correlations.

As our focus is on the entanglement between the detector flavor state and the propagation (source) degrees of freedom, we instead work with the reduced density matrix $\rho^{\mu}_{{e\tau}}$, obtained by tracing out the detected flavor $\nu_{\mu}$ and the subscript explicitly denotes the remaining flavor components. This concise description effectively encodes the necessary information to quantify predictability within the triality framework.
A detailed derivation of the reduced density matrix and the corresponding expression for $\mathcal{P}^2$ is presented in Appendices~\ref{RDM}, \ref{RDMnu}, and \ref{appendP}.

Choosing detector state to be $\nu_{\mu}$, the three-flavor system is reduced to a bipartite framework with n=2. Hence, in this framework, we consider the bipartite reduced density matrices $\rho_{{e\tau}}^{\mu}$, $\rho_{{\mu\tau}}^{\mu}$, $\rho_{{\mu e}}^{\mu}$. The diagonal elements of the time-evolved reduced density matrix can be expressed directly in terms of the measurable oscillation probabilities.

As discussed previously, tracing out flavor mode introduces mixedness in the reduced states, and it is precisely this mixedness that is captured by the predictability--visibility--entanglement triality relation.
The diagonal elements of the reduced density matrix, $\rho^{\mu}_{e\tau}$, can be written directly in terms of the measurable oscillation probabilities,
\begin{eqnarray}
    \rho_{e\tau_{_{11}}}^{\mu}  = P_{\mu\mu}~, \quad
\rho_{e\tau_{_{22}}}^{\mu}  = P_{\mu e}~, \quad
\rho_{e\tau_{_{33}}}^{\mu}  = P_{\mu\tau}~,
%\nonumber\\
\label{diagRDM}
\end{eqnarray}
where the normalization condition $P_{\mu\mu} + P_{\mu e} + P_{\mu\tau}=1$ holds at all times. Using these elements in the general definition of predictability, Eq.~\eqref{eq:predictability}, we obtain for the three-flavor case
\begin{equation}
    \mathcal{P}^{2}
    = 2\left(P_{\mu\mu}^{2} + P_{\mu e}^{2} + P_{\mu\tau}^{2}\right) - 1~.
    \label{predictability}
\end{equation}
This provides a direct operational measure of how strongly one flavor outcome dominates over the others. Predictability grows when a single transition probability becomes large and decreases when the flavor distribution becomes more democratic due to interference among the propagation amplitudes.

To make contact with the familiar two-flavor picture, consider the limit in which only $\nu_\mu$ and $\nu_e$ participate in oscillations. In this case, Eq.~(\ref{eq:predictability}) reduces to
\begin{eqnarray}
\mathcal{P}^{2}
    &= 2\left(P_{\mu\mu}^{2} + P_{\mu e}^{2}\right) - 1 = \left(P_{\mu\mu} - P_{\mu e}\right)^{2},\\ \nonumber
\label{eq:def-predictability}
\end{eqnarray}
so that $\mathcal{P} \propto |P_{\alpha\alpha} - P_{\alpha\beta}|$.
This expression highlights the intuitive interpretation of $\mathcal{P}^2$, which denotes the flavor imbalance within the two-flavor system. It also quantifies how distinguishable the flavor paths are.

Within the overall triality relation introduced earlier, $\mathcal{P}^2$ represents the particle-like contribution associated with the detector flavor. Its behavior therefore reflects the degree to which the detected neutrino maintains its identity during propagation, in competition with the wave-like visibility and the nonlocal correlation term arising from entanglement with the remaining flavor modes.
%%%%%%%%%%%%%%%%%%%%%%%%%%%%%%%%%%%%%%%  
\item \pmb{Visibility ($\mathcal{V}$) :} Visibility characterizes the strength of quantum interference among the flavor modes and captures the wave-like aspect of neutrino oscillations. In density-matrix language, visibility is governed by the off-diagonal elements of $\rho$, which encode the coherent superpositions responsible for flavor change. Following Refs.~\cite{Roy:2021wcg,Bittencourt:2023asd}, visibility for an $n$-level system is defined as
\begin{equation}
    \mathcal{V}^{2}
    = \frac{n}{n-1} \sum_{i\neq j} |\rho_{ij}|^{2},
    \label{eq:def-visibility}
\end{equation}
where $|\rho_{ij}|$ denotes the Hilbert--Schmidt coherence. 
In optical systems, the visibility $\mathcal{V}^2$ of interference fringes is defined in terms of the intensity contrast of the observed fringe pattern. When this concept is extended to the quantum domain, particularly for photons, the visibility is naturally expressed through the coherence of the quantum state, quantified by the off-diagonal density-matrix elements $|\rho_{ij}|$. This correspondence highlights that nonzero visibility arises only when the superposed states are coherent. In other words, the presence of coherent superposition is a necessary condition for observing interference effects or the wave nature of photons.
$\mathcal{V}^2$ quantity vanishes when the state loses all phase coherence and reaches its maximum value when interference among the modes is strongest.

In close analogy with the case of $\mathcal{P}^2$, the density matrix $\rho^{\mu}_{e\mu\tau}(t)$ given in Eq.~(\ref{eq:DM}) is not suitable for determining the visibility when detector-state entanglement is taken into account. Since $\rho^{\mu}_{e\mu\tau}(t)$ represents a pure flavor state, it does not retain the coherence information associated with correlations between the detector and propagation degrees of freedom. Consequently, it cannot adequately characterize the wave-like contribution encoded in the visibility.
When the detector is sensitive to the $\nu_\mu$ flavor, the three-flavor system reduces to an effective bipartite setting in which the detected mode $\nu_\mu$ shares a coherence with the complementary subsystem $\{\nu_e,\nu_\tau\}$. As discussed earlier, the total coherence of the detected mode is then given by the sum of its pairwise coherences with the other flavors, which is embedded in the reduced density matrices $\rho^{\mu}_{\mu e}$ and $\rho^{\mu}_{\mu\tau}$ (see Appendix~\ref{RDMnu}, \ref{appendV}). The off-diagonal elements of the reduced density matrices $\rho^{\mu}_{\mu e}$ and $\rho^{\mu}_{\mu\tau}$ quantify the Hilbert-Schmidt coherence, corresponding to the coherence between the flavor pairs ${\nu_{\mu}, \nu_e}$ and ${\nu_{\mu}, \nu_\tau}$, respectively. These terms capture the quantum coherence arising from the superposition of the associated flavor states and therefore directly encode the wave-like character of the system.
Using the off-diagonal elements of the reduce density matrix $\rho^{\mu}_{\mu e}$ and $\rho^{\mu}_{\mu\tau}$ lead to the expression
\begin{equation}
    \mathcal{V} = 2\left(P_{\mu\mu} P_{\mu e} + P_{\mu\mu} P_{\mu\tau}\right),
\label{eq:visibility}
\end{equation}
This provides a direct link between visibility and the measurable transition probabilities. This form makes clear that visibility is largest when the detected flavor $\nu_\mu$ simultaneously interferes with both $\nu_e$ and $\nu_\tau$, and decreases as the coherence between these channels is washed out during propagation. 

It is important to note that $\mathcal{P}^2$ and $\mathcal{V}^2$ are complementary quantities. Physically, this means that gaining more particle-like information about the neutrino flavor (higher $\mathcal{P}^2$) necessarily reduces the coherence responsible for interference (lower $\mathcal{V}^2$), highlighting the wave---particle dual nature of the system.

However, from Eq.~(\ref{eq:def-predictability}) and Eq.~(\ref{eq:visibility}) it can be readily seen that $\mathcal{P}^{2}+\mathcal{V}^{2}\neq 1$, which disrupts the saturation of duality relation. This suggests that predictability and visibility are connected through entanglement, and the duality relation can be expressed as a triality relation between predictability, visibility, and I-concurrence, which serves as a measure of entanglement.
%%%%%%%%%%%%%%%%%%%%%%%%%%%%%%%%%%%%%%%%  
\item  \pmb{Entanglement $\mathcal{E}$ :} 
Before analyzing the neutrino system in detail, it is instructive to recall how correlations arise in optical interferometry. Consider an interferometer equipped with a path-detecting device. A photon that may propagate along $n$ possible paths is then described by the state \cite{Roy:2021wcg, Roy:2021wcg}
\begin{equation}
    \ket{\psi}
    = \sum_{i=1}^{n} c_{i}\,
      \ket{\psi_{i}}\ket{d_{i}},
\end{equation}
where $\ket{\psi_{i}}$ denotes the photon's state associated with the $i^{\text{th}}$ path and $\ket{d_{i}}$ is the corresponding state of the path detector. The correlations between the photon and the path detector encode the extent to which path information is available, and they directly influence the observed interference pattern. Larger correlations imply stronger path information and, therefore, reduced visibility.

Motivated by this analogy, we now turn to the role of entanglement in three-flavor neutrino oscillations. As shown earlier, once a detector flavor (here $\nu_\mu$) is specified, the neutrino state naturally decomposes into the detected mode and the remaining two flavor mode subsystem. The corresponding predictability $\mathcal{P}^2$ and visibility $\mathcal{V}^2$ capture the particle-like and wave-like features of neutrinos. 
%To probe the third component that satisfies the triality relation, we derive the parameter which is equivalent to $1-\mathcal{P}^2-\mathcal{V}^2$.
%The remaining contribution in the triality relation,
%\begin{equation}
 %   \mathcal{P}^{2} + \mathcal{V}^{2} + \mathcal{E}^{2} = 1,
  %  \label{triality}
%\end{equation}
%is provided by the correlation term $\mathcal{E}^2$, which quantifies the nonlocal correlations shared between the detected flavor and the other two flavor modes.

To characterize the contribution from the entanglement, we employ the I-concurrence \cite{Bhaskara:2017uzs}, which is an entanglement measure arising when one subsystem is traced out of a pure global state. For an $n$-level system, a normalized entanglement measure is defined as
\begin{equation}
    \mathcal{E}^{2}
    = \frac{2}{n(n-1)}\, E^2,
\end{equation}
\begin{table}[] 
\centering 
\small 
{ 
\hspace{0cm} 
\begin{tabular}{|c|c|} 
\hline \multicolumn{2}{|c|}{Standard oscillation parameters} 
\\ \hline \hline Parameter & Best-fit value \\ & NO (IO) \\ \hline \hline 
$\theta_{12}$ & $33.68^{\circ} (33.68^{\circ})$ \\ 
\hline $\theta_{13}$ & $8.52^{\circ} (8.58^{\circ})$ \\ \hline $\theta_{23}$ & $48.5^{\circ}(48.6^\circ)$ \\ \hline $\Delta m_{21}^2 (\rm eV^2)$ & $7.49 \times 10^{-5} (7.49 \times 10^{-5})$\\ 
\hline $\Delta m_{31}^2 (\rm eV^2)$ & $2.534 \times 10^{-3}(-2.484 \times 10^{-3})$ \\ \hline \end{tabular} \caption{Best-fit and 3$\sigma$ ranges of standard oscillation parameters \cite{Esteban:2024eli}} \label{oscparams} } 
\end{table}
where $E$ denotes the I-concurrence and reduces to the usual concurrence for a bipartite system. Applying this framework to the three-flavor neutrino system with $\nu_\mu$ as the detector state, we find that the total correlation of $\nu_\mu$ with the remaining subsystem $\{\nu_e,\nu_\tau\}$ decomposes into a sum of its pairwise bipartite correlations $(C_{e|\mu\tau}^{2} + C_{\tau|\mu e}^{2})$.
\begin{eqnarray}
   E^2 &=& C_{e|\mu\tau}^{2} + C_{\tau|\mu e}^{2} \nonumber \\
    &=&4 P_{\mu\mu}P_{\mu e}+4P_{\mu\mu}P_{\mu\tau}+8P_{\mu e}P_{\mu\tau}~,
\label{eq:entanglement}
\end{eqnarray}
%\begin{eqnarray}
 %   \mathcal{E}^{2}
  %      &= \frac{1}{2}\left(E_{\mu e}^{2} + E_{\mu\tau}^{2}\right)
   %      = \frac{1}{2}\left(C_{e|\mu\tau}^{2} + C_{\tau|\mu e}^{2}\right),
   % \label{eq:entanglement}
%\end{eqnarray}
where $C_{e|\mu\tau}$ and $C_{\tau|\mu e}$ denote the concurrences describing the bipartite entanglement between $\nu_e$ and $\nu_e$, and between $\nu_\mu$ and $\nu_\tau$, respectively (see Appendix~\ref{appendE}). Eq.~(\ref{eq:entanglement}) thus reveals how the detector flavor shares quantum correlations with the remaining propagation modes.
\end{itemize}
Combining the expressions for predictability, visibility, and entanglement, we recover the exact triality condition,
\begin{equation}
    \mathcal{E}^{2} + \mathcal{P}^{2} + \mathcal{V}^{2} = 1.
\end{equation}
This establishes that three-flavor neutrino oscillations naturally satisfy a complementarity relation in which the loss of local flavor information (predictability and visibility) is precisely balanced by the correlations generated with the undetected flavor subspace. Such a structure does not arise in a two-flavor description, since tracing out the detector flavor leaves only a single remaining degree of freedom, sufficient to saturate the duality relation between $\mathcal{P}^{2}$ and $\mathcal{V}^{2}$ without requiring an additional correlation term.

In the next section, we apply this formalism to two long-baseline accelerator experiments, DUNE and T2K, which probe different regions of parameter space due to their distinct baselines and energies.
%%%%%%%%%%%%%%%%%%%%%%%%%%%%%%%%%%%%%%%%%%%%%%%%%%
\section{Experimental Implications: DUNE and T2K}
\label{experiment}
%%%%%%%%%%%%%%%%%%%%%%%%%%%%%%%%%%%%%%%%%%%%%%%%%%
We now examine the dynamical evolution of the quantum triality relation in realistic experimental settings, focusing on the long-baseline experiments \textsf{DUNE}~\cite{DUNE:2020jqi} and \textsf{T2K}. All oscillation probabilities computations are obtained using the \textsf{GLoBES-v3.0}  framework~\cite{Huber:2004ka, Huber:2007ji}. The two experiments provide two different baselines---1300 km for \textsf{DUNE} and 295 km for \textsf{T2K}---allowing us to probe the triality structure across distinct oscillation regimes to observe the effect of both vacuum (\textsf{T2K}) and matter effect (\textsf{DUNE}). The oscillation parameters employed in our simulation follow the \textsf{NuFIT}~v6.0 global analysis~\cite{Esteban:2024eli}, summarized in Table~\ref{oscparams}.
Figures~\ref{DUNE_CCR} and \ref{T2K_CCR} display the behavior of the triality components as functions of neutrino energy. The upper panels show the three complementary quantities, $\mathcal{P}^{2}$ (solid orange curve), $\mathcal{V}^{2}$ (solid blue curve), and $\mathcal{E}^{2}$ (solid green curve). In contrast, the lower panels present the corresponding appearance and disappearance probabilities, $P_{\mu\mu}$ (dashed orange curve), $P_{\mu e}$ (dashed green curve), and $P_{\mu\tau}$ (dashed blue curve).
\subsection{Analysis for \textsf{DUNE}}
\label{DUNE}
\begin{figure}[hbt!]
  	\centering
  	\hspace{-0.11cm}
    \vspace*{-0.55cm}
  	%\includegraphics[width=0.35\textwidth]{sRHMD_final.pdf}
  	%\hspace{-0.6cm}
  	\includegraphics[width=0.45\textwidth]{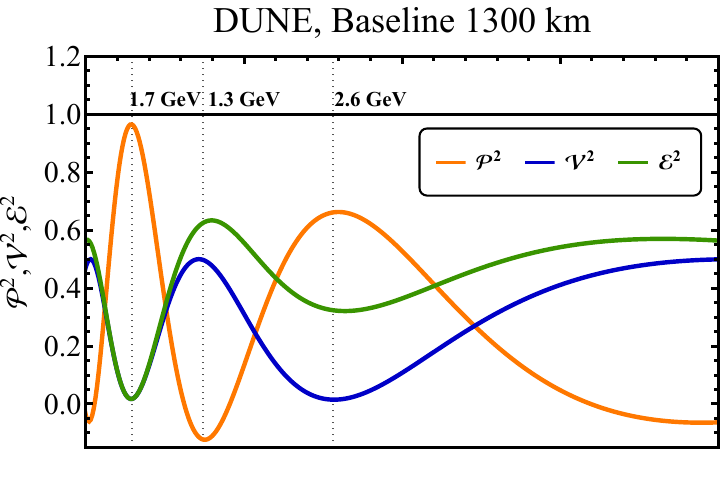}
    \hspace{0.2cm}
   %  \vspace*{-1.2cm}
  	%\includegraphics[width=0.35\textwidth]{sRHMDt_final.pdf}
  	\includegraphics[width=0.457\textwidth]{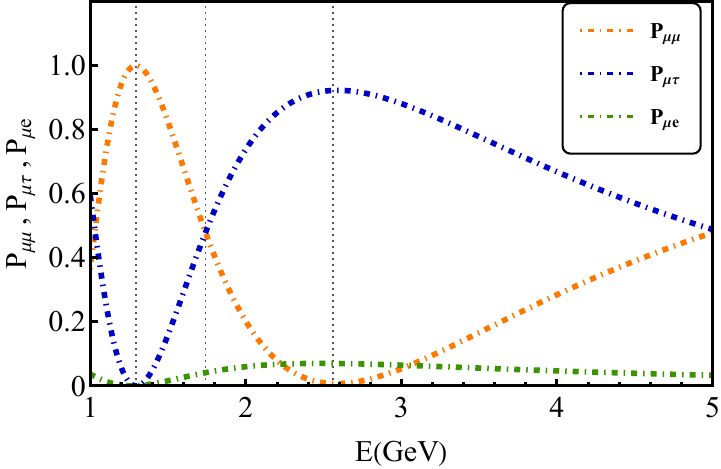}
  	\caption{Triality condition in the three-flavor case at a \textsf{DUNE} baseline. Oscillation dynamics generate regimes of high visibility (wave-like behavior), strong predictability (particle-like behavior), and maximal entanglement (mixed states). The interplay among the three quantities demonstrates flavor monogamy,i.e., whenever two flavors are maximally entangled, the third is excluded. The residual plot verifies $\mathcal{P}^2 + \mathcal{V}^2 + \mathcal{E}^2 = 1$ across the full energy range.} 
  	\label{DUNE_CCR}
  \end{figure}

%\textbf{DUNE:} 
Figure~\ref{DUNE_CCR} illustrates how the triality relation manifests across the \textsf{DUNE} energy range. In particular, near the first oscillation maximum ($  E\simeq 2.6$~GeV), the triality condition is satisfied through a nontrivial redistribution among predictability, visibility, and entanglement. To highlight the underlying physics, we analyze three representative energy bins:

\begin{itemize}
    \item 1.3~GeV, where predictability approaches a global maximum,
    \item 1.7~GeV, where predictability reaches a minimum and coherence is largest,
    \item 2.6~GeV, corresponding to the first oscillation maximum of $P_{\mu e}$, where \textsf{DUNE} has maximal sensitivity.
\end{itemize}
  \begin{table*}[hbt!]
\centering
\renewcommand{\arraystretch}{1.3} % increases row height
\setlength{\tabcolsep}{12pt} % adds column spacing
{\large
\begin{tabular}{c c c c c c c c}
\hline\hline
Experiment & $E$ (GeV) & $\mathcal{P}^{2}$ & $\mathcal{V}^{2}$ & $\mathcal{E}^{2}$ & $P_{\mu\mu}$ & $P_{\mu e}$ & $P_{\mu\tau}$ \\
\hline
\multirow{3}{*}{DUNE} 
 & 2.6 & 0.7 & 0 & 0.3 & 0.01  & 0.07 & 0.92 \\
 & 1.7 & 0 & 0.4 & 0.6 & 0.47  & 0.06 & 0.47 \\
 & 1.3 & 1 & 0 & 0 & 1.0 & 0 & 0 \\
\hline
\multirow{2}{*}{T2K} 
 & 0.6 & 0.7 & 0 & 0.3 & 0 & 0.05 & 0.95 \\
 & 1.2 & 0 & 0.4 & 0.5 & 0.48  & 0.02 & 0.48 \\
\hline\hline
\end{tabular}
}
\caption{Triality condition parameters and oscillation probabilities for \textsf{DUNE} and \textsf{T2K} at characteristic energies.}
\label{tab:triality}
\end{table*}
\textbf{1.3~GeV Energy Bin:}  
At $1.3$~GeV, the system is dominated by the survival channel, with $P_{\mu\mu}\approx 1$ and $P_{\mu e}, P_{\mu\tau}\approx 0$. Consequently, $\mathcal{P}^{2}$ approaches unity while both $\mathcal{V}^{2}$ and $\mathcal{E}^{2}$ vanish. This energy region, therefore, describes a nearly separable state of $\nu_\mu$, exhibiting an almost purely particle-like behavior. The absence of coherence and entanglement is consistent with the dominance of a single propagation path.
\medskip
\noindent  
\\ \textbf{1.7~GeV Energy Bin:}  
In contrast, at $1.7$~GeV the system approaches a maximally mixed regime. Here, $\mathcal{P}^{2}\approx 0$, while $\mathcal{V}^{2}\approx 0.4$ and $\mathcal{E}^{2}\approx 0.6$ attain their largest values. The lower panel shows that all three probabilities are nonzero, with $P_{\mu\mu}\approx P_{\mu\tau}\approx 0.5$. The large value of $\mathcal{E}^{2}$ indicates strong entanglement between the detector state and the propagation modes, which suppresses predictability. The simultaneous enhancement of visibility reflects the pronounced wave-like behavior in this region, although entanglement partially mitigates the achievable coherence.

\noindent
\\ \textbf{2.6~GeV Energy Bin:}  
At the first oscillation maximum, the system exhibits a balanced interplay between particle-like and wave-like behavior. Predictability attains a local maximum ($\mathcal{P}^{2}\approx 0.7$), while the entanglement term decreases to $\mathcal{E}^{2}\approx 0.3$. The visibility is minimal, indicating that the wave nature is not solely captured by $\mathcal{V}^{2}$ when the state is mixed. The nonzero values of both $\mathcal{P}^{2}$ and $\mathcal{E}^{2}$ show that information about wave-like behavior is distributed between the visibility and entanglement. 
The inability of $\mathcal{P}^{2}$ to reach unity demonstrates that entanglement between the detected and propagated flavor modes limits the purity of the detected state. This behavior reflects the intrinsic quantum correlations present in three-flavor neutrino oscillations. Interestingly, at this particular energy bin, both the particle-like and
wave-like aspects of neutrino behavior can be accessed simultaneously.
While the predictability characterizes the particle nature
$\mathcal{P}^2$, the wave nature is revealed through the entanglement
measure $\mathcal{E}^2$, allowing both features to be probed in the same
regime.
%%%%%%%%%%%%%%%%%%%%%%%%%%
%\section{Verifying triality relation with DUNE and T2K}
\subsection{Analysis for T2K} 
\label{T2K}
\begin{figure}[hbt!]
	\centering
	\hspace{-0.24cm}
	\vspace*{-0.55cm} 	%\includegraphics[width=0.35\textwidth]{sRHMD_final.pdf}
	%\hspace{-0.6cm}
	\includegraphics[width=0.45\textwidth]{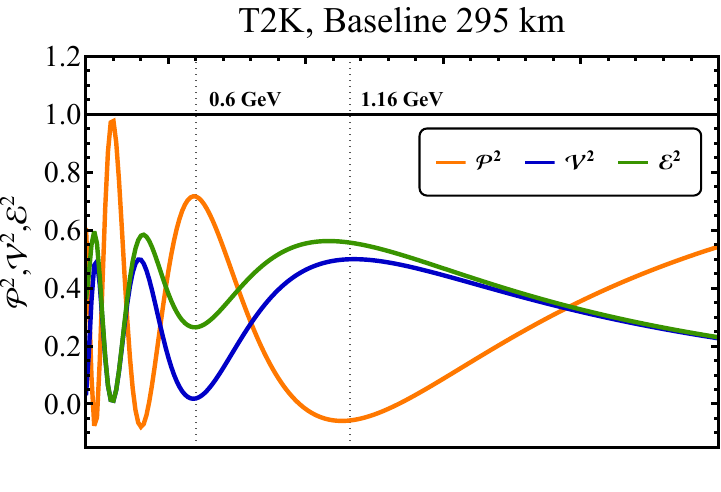}
	\hspace{0.2cm}  	%\includegraphics[width=0.35\textwidth]{sRHMDt_final.pdf}
	\includegraphics[width=0.465\textwidth]{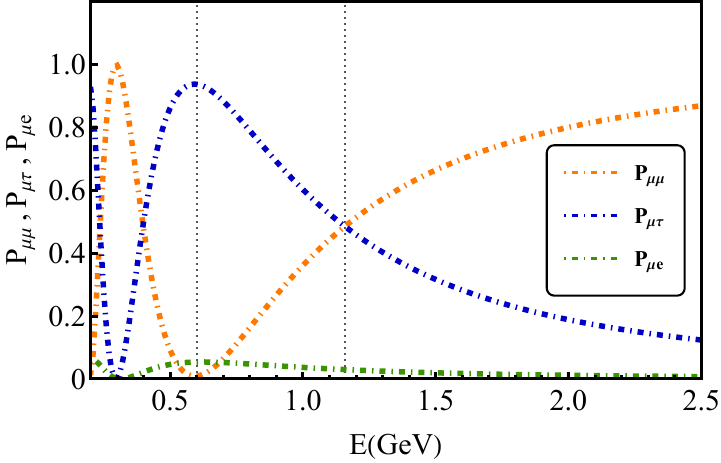}
	\caption{Three-flavor neutrino oscillations as a natural realization of triality. The visibility $\mathcal{V}^{2}$, predictability $\mathcal{P}^{2}$, and entanglement $\mathcal{E}^{2}$ are shown as functions of neutrino energy for \textsf{T2K} with baseline of 295 km. The three quantities exchange dominance across the energy spectrum while preserving the exact triality relation. Entanglement monogamy emerges, with two flavors maximally correlated while the third decouples.} 
	\label{T2K_CCR}
\end{figure}

%\textbf{T2K:} 
Figure~\ref{T2K_CCR} illustrates the behavior of the triality components for the \textsf{T2K} experiment. 
Due to shorter baseline length, $\textsf{T2K}$ mainly manifests the vacuum effect on the triality relation of neutrino.
To examine how the complementarity relation manifests in this setup, we focus on two representative energy bins:
\begin{itemize}
    \item $0.6$~GeV, corresponding to the first oscillation maximum of $P_{\mu e}$, where T2K achieves its highest sensitivity,
    \item $1.16$~GeV, where the predictability $\mathcal{P}^{2}$ reaches a pronounced minimum.
\end{itemize}
\textbf{0.6~GeV Energy Bin:}  
At $0.6$~GeV, T2K probes the first oscillation maximum, a region characterized by significant flavor conversion. Here, predictability is moderately enhanced, with $\mathcal{P}^{2}\approx 0.7$, indicating a partially particle-like behavior of the detected flavor. Visibility is strongly suppressed, $\mathcal{V}^{2}\approx 0$, implying that the interference component alone does not fully capture the wave-like features of the state. At the same time, the entanglement term reaches a relatively large value, $\mathcal{E}^{2}\approx 0.3$, reflecting substantial entanglement between the detected $\nu_\mu$ mode and the remaining flavor subspace.
The combined behavior of these quantities closely resembles that observed at the first oscillation maximum in \textsf{DUNE}. The neutrino exhibits both particle-like and wave-like character, but the wave component is distributed between the visibility and the entanglement contributions. In other words, the mixed nature of the state prevents visibility from capturing all interference effects, and part of the underlying wave information is encoded in the entanglement represented by $\mathcal{E}^{2}$.
\medskip
\noindent
\\ \textbf{1.16~GeV Energy Bin:}  
At $1.16$~GeV, the triality structure in T2K displays behavior characteristic of a strongly mixed quantum state. In this region, predictability reaches a pronounced minimum, $\mathcal{P}^{2}\approx 0$, indicating that the detected flavor retains almost no particle-like distinguishability. Correspondingly, both visibility and entanglement attain comparatively large values, with $\mathcal{V}^{2}\approx 0.45$ and $\mathcal{E}^{2}\approx 0.55$, which together saturate the triality condition. 
From the probability distribution, all three channels---$P_{\mu\mu}$, $P_{\mu e}$, and $P_{\mu\tau}$---are nonzero, reflecting a broad sharing of flavor amplitudes among the available modes. This redistribution is precisely what suppresses predictability while simultaneously enhancing wave-like and correlation-based features. The sizable value of $\mathcal{E}^{2}$ indicates that the detector flavor $\nu_\mu$ becomes strongly entangled with the remaining propagation subspace, reducing the amount of information recoverable from the detected flavor mode.
Thus, the $1.16$~GeV bin highlights a regime where the neutrino behaves predominantly as a delocalized quantum object, where $\mathcal{V}^2$ captures the coherent interference among flavor channels. At the same time, $\mathcal{E}^2$ accounts for the quantum correlations generated through the mixing process. Together, these contributions illustrate the complete complementarity structure intrinsic to three-flavor oscillations. 

By contrast, at $E\simeq 1.16\ \mathrm{GeV}$ the system enters a coherence-dominated regime where predictability vanishes ($\mathcal{P}^{2}\simeq 0$) due to the nearly equal partition between survival and $\nu_{\tau}$ probabilities ($P_{\mu\mu}\simeq P_{\mu\tau}\simeq 0.48$),
Visibility is large ($\mathcal{P}^2\simeq 0.79$) signaling substantial interference among flavor amplitudes, and entanglement is small but nonzero ($\mathcal{E}^2\simeq 0.21$), consistent with the minor $\nu_{e}$ channel. 
%These results show that at \textsf{T2K} energies and baseline, where matter effects are modest, the triality condition is realized through a shifting balance conversion at the oscillation maximum, which enhances entanglement, whereas at higher energies, coherence dominates with reduced three-flavor correlations.
\\ Together, these results show that at the shorter \textsf{T2K} baseline the oscillation pattern remains essentially vacuum-like. In contrast, the longer \textsf{DUNE} baseline exhibits the expected matter-induced distortions of the oscillation probabilities. However, the triality components themselves---$\mathcal{P}^2$, $\mathcal{v}^2$, and $\mathcal{E}^2$---do not experience any significant enhancement or suppression due to matter effects. The plots clearly indicate that although matter modifies the oscillation probabilities, it does not alter the fundamental balance among the three quantities. As a result, the triality relation continues to hold with equal exactness in both vacuum-dominated and matter-dominated regimes, demonstrating its robustness as a structural feature of three-flavor quantum evolution.
\section{Conclusion}
\label{conclusion}
In this work, we presented a quantum-information perspective on three-flavor neutrino oscillations by examining the exact triality relation that links predictability ($\mathcal{P}^2$), visibility ($\mathcal{V}^2$), and entanglement ($\mathcal{E}^2$). Treating the detected flavor as an effective subsystem shows that, although the global neutrino state remains pure throughout propagation, the reduced flavor states inevitably acquire mixedness and correlations once a specific detection channel is specified.
The wave-like information carried by the neutrino is therefore not captured solely by the visibility term, $\mathcal{V}^2$. Part of this information is encoded in the coherence among the propagating flavor amplitudes. At the same time, the remainder is carried by the entanglement, $\mathcal{E}^2$, formed between the detector state and the rest of the flavor space. In this work, we quantified this entanglement using the I-concurrence, which offers a natural measure of bipartite correlations in the three-flavor framework.
This viewpoint broadens the familiar wave-particle duality by showing that the informational content of flavor oscillations is shared across all three quantities: $\mathcal{P}^2$, $\mathcal{V}^2$, and $\mathcal{E}^2$. Together, they reveal a richer and more complete complementarity structure intrinsic to three-flavor neutrino evolution.
Our analysis also clarifies why this triality relation is unique to the three-flavor case. A two-flavor system behaves effectively as a bipartite configuration, and tracing over the detected flavor leaves a single remaining degree of freedom. In such a situation, the triality relation reduces to the standard duality relation between predictability and visibility. Although entanglement can still exist between different flavor modes, the bipartite structure in two-flavor neutrino oscillation is insufficient to support the additional correlation term required for triality. In contrast, the three-flavor system naturally accommodates entanglement between the detector flavor and the remaining two-mode propagation subspace, enabling the complete triality framework to emerge.

Our analysis explored the energy-dependent behavior of the triality components using realistic simulations performed with \textsf{GLoBES} for two long-baseline neutrino experiments, \textsf{DUNE} and \textsf{T2K} (Fig.~\ref{DUNE_CCR},\ref{T2K_CCR}), which probe different oscillation regimes due to their distinct baselines. This comparison enabled us to study the triality relation in both environments that are effectively vacuum-dominated (\textsf{T2K}) and those where matter effects play a significant role (\textsf{DUNE}). In both cases, the triality condition remained valid across the entire energy range, demonstrating that the balance among predictability, visibility, and entanglement is a robust feature of three-flavor evolution, independent of whether the propagation occurs in vacuum or matter.

The variations in $\mathcal{P}^2$ (solid orange line), $\mathcal{V}^2$ (solid blue line), and $\mathcal{E}^2$ (solid green line) with energy exhibit clear oscillatory behavior, reflecting the continuous exchange between particle-like and wave-like characteristics of the neutrino. For \textsf{DUNE}, we probed three representative energies (1.3, 1.7, and 2.6~GeV), while for \textsf{T2K} we examined two key energies (0.6 and 1.2~GeV), each including the first oscillation maximum of $P_{\mu e}$. These benchmarks illustrate three distinct regimes. When $\mathcal{P}^2$ approaches unity, both $\mathcal{V}^2$ and $\mathcal{E}^2$ vanish, indicating a highly separable state dominated by the survival channel, with $P_{\mu\mu}\approx 1$ and negligible appearance probabilities. At the opposite extreme, when $\mathcal{P}^2$ reaches its minimum, both $\mathcal{V}^2$ and $\mathcal{E}^2$ attain large values, revealing a strongly mixed state exhibiting pronounced wave-like behavior and significant quantum correlations.

At the first oscillation maximum of $P_{\mu e}$, the three quantities share the information content in a nontrivial manner: predictability recovers partially, visibility is suppressed, and entanglement carries a significant portion of the wave-like information. This feature underscores an essential aspect of three-flavor oscillations—coherence lost at the level of visibility can reappear within the entanglement term, which reflects the nonlocal correlations between the detector flavor and the remaining propagation modes. The numerical values corresponding to these energy bins are summarized in Table~\ref{tab:triality}, reaffirming the quantitative fulfillment of the triality condition in both experiments.

Overall, our study demonstrates that entanglement forms a natural bridge between the wave-like and particle-like aspects of neutrino behavior. By framing flavor transitions within a quantum information perspective, the triality relation provides a unified and intuitive description that links measurable oscillation probabilities to the underlying quantum structure. This framework offers a promising foundation for future precision studies in neutrino physics and for exploring broader questions in quantum foundations. 
Although the particle-like and wave-like aspects of neutrino behavior
cannot be probed simultaneously using the standard quantities
$\mathcal{P}^2$ and $\mathcal{V}^2$, because of their inherent
complementarity, our analysis reveals that this limitation can be
overcome by incorporating entanglement. In particular, for three-flavor
neutrino oscillations, we find that at the first oscillation maximum of
both experiments, the combined information carried by the predictability
of the detected flavor state and the entanglement measure makes it
possible to access the particle and wave nature of neutrinos at the same
time. A natural extension of this work would involve repeating the triality analysis with the detector state chosen as $\nu_e$, which would offer complementary insight into appearance channels and the structure of flavor correlations in upcoming long-baseline experiments.

\label{sec5}
%%

%%%%%%%%
\section{Acknowledgement}
%%%%%%%%
Rajrupa Banerjee would like to thank the Ministry of
Electronics and IT for the financial support through the Visvesvaraya fellowship scheme for carrying out this research work. SP acknowledges the financial support under MTR/2023/000687 funded by SERB, Govt. of India. 
%%%%%%%%%%
\appendix
\section{Derivation of the Reduced Density Matrix for a Three-Qubit State}
\label{RDM}
In quantum systems with multiple degrees of freedom, the density matrix provides a complete and 
meaningful description of both pure and mixed states, allowing one to systematically characterize coherence, 
correlations, and entanglement. In particular, reduced density matrices—obtained by tracing out unobserved 
subsystems—play a central role in quantifying how wave-like interference, particle-like distinguishability, 
and entanglement coexist, thereby forming the basis of the quantum triality relation. 
This Appendix is devoted to the explicit realization of the general formalism for three-flavor neutrino oscillations, 
wherein the flavor basis provides a natural and physically transparent mapping to a three-qubit Hilbert space. 
We fix the computational basis for a three-qubit Hilbert space 
$\mathcal{H}=\mathcal{H}_A\otimes\mathcal{H}_B\otimes\mathcal{H}_C$,
with each single-qubit space $\mathcal{H}_{\alpha}$ ($\alpha=A,B,C$)
spanned by the orthonormal basis $\{\ket{0},\ket{1}\}$. 
The resulting product basis is chosen in the standard ordering
\begin{equation}
\Big\{
\ket{000}, \ket{001}, \ket{010}, \ket{011},
\ket{100}, \ket{101}, \ket{110}, \ket{111}
\Big\}, \nonumber 
\end{equation}
where $\ket{ijk}\equiv\ket{i}_A\otimes\ket{j}_B\otimes\ket{k}_C$ with
$i,j,k\in\{0,1\}$.

An arbitrary pure state of the three-qubit system can then be expressed
as a coherent superposition of these basis states,
\begin{equation}
\ket{\Psi}
=
\sum_{i,j,k\in\{0,1\}} c_{ijk}\,\ket{ijk},
\qquad
\sum_{i,j,k}|c_{ijk}|^2 = 1,
\label{eq:psi3}
\end{equation}
where $c_{ijk}\in\mathbb{C}$ are complex probability amplitudes and the
normalization condition ensures $\langle\Psi|\Psi\rangle=1$.

The density operator associated with this pure state is defined as
%\begin{equation}
$\rho \equiv \ket{\Psi}\bra{\Psi}$.
%\end{equation}
In the computational basis, the matrix elements of $\rho$ are given by
\begin{equation}
\rho_{(ijk),(pqr)}
\equiv
\bra{ijk}\rho\ket{pqr}
=
c_{ijk}\,c^{*}_{pqr},
\end{equation}
where $(ijk)$ and $(pqr)$ label the basis states of the composite system
and the superscript $^{*}$ denotes complex conjugation.
Consequently, the density matrix $\rho$ is an $8\times8$ Hermitian,
positive semi-definite matrix with $\mathrm{Tr}(\rho)=1$ and
$\rho^2=\rho$, indicating that it has rank one and therefore represents
a pure three-qubit state.
\subsection{Reduced density matrices via partial trace}
To analyze correlations and entanglement among subsystems, we consider reduced density matrices obtained by tracing out one or more degrees of freedom.
\medskip
\noindent
\\ {\bf Single-qubit reduced density matrix:}\,
The reduced density matrix of qubit \(A\) is obtained by tracing out qubits \(B\) and \(C\),
\begin{equation}
\rho_A = \mathrm{Tr}_{BC}(\rho)
= \sum_{j,k=0}^{1} \bra{jk}_{BC}\,\rho\,\ket{jk}_{BC}.
\label{singleRDM}
\end{equation}
Evaluating the trace explicitly yields
\begin{equation}
\rho_A =
\begin{pmatrix}
\sum_{j,k}|c_{0jk}|^2 & \sum_{j,k} c_{0jk} c_{1jk}^{*} \\[4pt]
\sum_{j,k} c_{1jk} c_{0jk}^{*} & \sum_{j,k}|c_{1jk}|^2
\end{pmatrix}.
\label{eq:rhoA}
\end{equation}
Analogous expressions for \(\rho_B\) and \(\rho_C\) follow by permutation of indices.

\noindent
{\bf Two-qubit reduced density matrix:}\, 
The reduced density matrix of qubits \(A\) and \(B\) is obtained by tracing out qubit \(C\),
\begin{equation}
\rho_{AB} = \mathrm{Tr}_{C}(\rho)
= \sum_{k=0}^{1} \bra{k}_{C}\,\rho\,\ket{k}_{C}.
\label{twoRDM}
\end{equation}
\subsection{Reduced density matrices for the W state}
As an explicit example, we consider the symmetric three-qubit W state,
\begin{equation}
\ket{W} = \frac{1}{\sqrt{3}}\big(\ket{001} + \ket{010} + \ket{100}\big).
\end{equation}
The only nonvanishing amplitudes are
\(c_{001} = c_{010} = c_{100} = 1/\sqrt{3}\).
\medskip
\noindent
\\{\bf Single-qubit reduced state:}\,
Substituting these coefficients into Eq.~\eqref{eq:rhoA}, the reduced density matrix for qubit \(A\) becomes
\begin{equation}
\rho_A =
\begin{pmatrix}
\frac{2}{3} & 0 \\
0 & \frac{1}{3}
\end{pmatrix}.
\end{equation}
By symmetry, the same reduced state holds for qubits \(B\) and \(C\). Each single-qubit reduced state is mixed, 
indicating entanglement between the qubit and the remaining subsystem.
\medskip
\noindent
\\{\bf Two-qubit reduced state:}\,
Tracing out qubit \(C\), the reduced density matrix \(\rho_{AB}\) is obtained in the ordered basis
\(\{\ket{00},\ket{01},\ket{10},\ket{11}\}_{AB}\),
\begin{equation}
\rho_{AB}
=
\frac{1}{3}
\begin{pmatrix}
1 & 0 & 0 & 0 \\
0 & 1 & 1 & 0 \\
0 & 1 & 1 & 0 \\
0 & 0 & 0 & 0
\end{pmatrix}.
\end{equation}
The presence of nonvanishing off-diagonal elements reflects bipartite coherence and entanglement between qubits \(A\) and \(B\).

\medskip
\noindent
These reduced density matrices generally describe mixed states, even though the global three-qubit state \(\rho\) is pure. The emergence of mixedness in the reduced description is a direct manifestation of quantum entanglement between the traced-out subsystem and the remaining degrees of freedom.

The off-diagonal elements of the reduced density matrices encode quantum coherence between different basis states, whereas the diagonal elements represent classical population probabilities. Consequently, reduced density matrices provide the appropriate framework for quantifying both wave-like properties, through coherence measures, and particle-like properties, through predictability.

In this way, the reduced density matrix formalism forms the foundation for analyzing bipartite and multipartite entanglement in three-qubit systems. Naturally, it connects to operational measures such as visibility, predictability, and entanglement within the triality framework.
\section{Reduced Density Matrices in Three-Flavor Neutrino Oscillation}
\label{RDMnu}
In the three-flavor framework, the time-evolved state of an initially produced muon neutrino $\ket{\nu_\mu(t)}$ occupies the three orthogonal flavor modes $(\nu_e,\nu_\mu,\nu_\tau)$ in the extended occupation-number basis, $ \ket{\nu_{e}} \equiv \ket{1}_{e}\otimes\ket{0}_{\mu}\otimes\ket{0}_{\tau}$, $ \ket{\nu_{\mu}} \equiv \ket{0}_{e}\otimes\ket{1}_{\mu}\otimes\ket{0}_{\tau}$ and $ \ket{\nu_{\tau}} \equiv \ket{0}_{e}\otimes\ket{0}_{\mu}\otimes\ket{1}_{\tau}$. 
Since a neutrino is a fermion, following Pauli's exclusion principle, only $\{\ket{100},\ket{010},\ket{001}\}$ is relevant for a neutrino. The choice of the basis for the three-flavor neutrino system leads to the formation of the general W state configuration of the time-evolved state vector $\nu_{\mu}(t)$. Substituting $\{A,~B,~C\}=\{\nu_e,~\nu_\mu,~\nu_\tau\}$, the construction of the W state for the neutrino can be treated as,
\begin{eqnarray}
    \ket{\nu_{\mu}(t)}=\Tilde{U}_{\mu e}\ket{100}+\Tilde{U}_{\mu\mu}\ket{010}+\Tilde{U}_{\mu \tau}\ket{001},
\label{eq:appenState}
\end{eqnarray}
To analyze the pattern of quantum correlations shared among these modes, we construct reduced density matrices by tracing over selected subsystems.

Starting from the full density operator $\rho^\mu_{e\mu\tau}(t)=\ket{\nu_\mu(t)}\bra{\nu_\mu(t)}$, the bipartite reduced density matrices for the $(e\mu)$, $(e\tau)$, and $(\mu\tau)$ subsystems are obtained by tracing over the remaining flavor mode. 
\medskip
\noindent
\\{\bf Single-qubit reduced state:}\,
Folloing Eq,~(\ref{singleRDM}) from Sec.~\ref{RDM} for an initial $\nu_\mu$ state, the matrices are given by
\begin{eqnarray}
    \rho_{e\mu}^{\mu} & =\begin{pmatrix}
          |\widetilde{U}_{\mu\tau}|^{2}  & 0 & 0 & 0\\
          0 & |\widetilde{U}_{\mu\mu}|^{2}  & \widetilde{U}_{\mu\mu}\widetilde{U}^{*}_{\mu e} & 0\\
          0 & \widetilde{U}_{\mu e}\widetilde{U}^{*}_{\mu\mu} & |\widetilde{U}_{\mu e}|^{2} & 0\\
          0 & 0 & 0 & 0
      \end{pmatrix},
      \nonumber\\[2mm]
    \rho_{e\tau}^{\mu} & =\begin{pmatrix}
          |\widetilde{U}_{\mu\mu}|^{2} & 0 & 0 & 0\\
          0 & |\widetilde{U}_{\mu\tau}|^{2} & \widetilde{U}_{\mu\tau}\widetilde{U}^{*}_{\mu e} & 0 \\
          0 & \widetilde{U}_{\mu e}\widetilde{U}^{*}_{\mu\tau} & |\widetilde{U}_{\mu e}|^{2} & 0\\
          0 & 0 & 0 & 0
      \end{pmatrix},
      \nonumber\\[2mm]
    \rho_{\mu\tau}^{\mu} & =\begin{pmatrix}
          |\widetilde{U}_{\mu e}|^{2} & 0 & 0 & 0 \\
          0 & |\widetilde{U}_{\mu\tau}|^{2} & \widetilde{U}_{\mu\tau}\widetilde{U}^{*}_{\mu\mu} & 0 \\
          0 & \widetilde{U}_{\mu\mu}\widetilde{U}^{*}_{\mu\tau} & |\widetilde{U}_{\mu\mu}|^{2} & 0 \\
          0 & 0 & 0 & 0
      \end{pmatrix}.
      \label{eq:reducedDM}
\end{eqnarray}
The diagonal elements reproduce the corresponding oscillation probabilities, while the off-diagonal terms encode the coherence between flavor modes.
\medskip
\noindent
\\{\bf Two-qubit reduced state:}\,
Tracing once more over one of the remaining subsystems yields the reduced single-flavor density matrices,
\begin{eqnarray}
    \rho^{\mu}_{e} & =\begin{pmatrix}
        |\widetilde{U}_{\mu\tau}|^{2}+|\widetilde{U}_{\mu\mu}|^{2} & 0\\
        0 & |\widetilde{U}_{\mu e}|^{2}
    \end{pmatrix},
    \nonumber\\[2mm]
    \rho^{\mu}_{\mu} & =\begin{pmatrix}
        |\widetilde{U}_{\mu\tau}|^{2}+|\widetilde{U}_{\mu e}|^{2} & 0\\
        0 & |\widetilde{U}_{\mu\mu}|^{2}
    \end{pmatrix},
    \nonumber\\[2mm]
    \rho^{\mu}_{\tau} &=\begin{pmatrix}
        |\widetilde{U}_{\mu e}|^{2}+|\widetilde{U}_{\mu\mu}|^{2} & 0\\
        0 & |\widetilde{U}_{\mu\tau}|^{2}
    \end{pmatrix}.
\label{reduceDMsingle}
\end{eqnarray}
These single-mode density matrices determine all measurable flavor observables. In particular, the von Neumann entropy constructed from $\rho^{\mu}_{\alpha}$ quantifies the degree of entanglement between a given flavor mode and the remaining modes, providing the basic ingredients for the entanglement measures discussed in subsequent sections.
\section{Calculation of Predictability ($\mathcal{P}^2$) in Three-Flavor Neutrino Oscillation}
\label{appendP}

To evaluate the predictability in the three-flavor neutrino system, we consider the initial state to be a muon neutrino, $\nu_\mu$, which is relevant for accelerator-based experiments such as DUNE. Treating $\nu_\mu$ as the detected flavor, we trace over this mode to obtain the reduced density matrix $\rho^{\mu}_{e\tau}$ describing the remaining $(\nu_e,\nu_\tau)$ subsystem. From Eq.~(\ref{eq:reducedDM}), the diagonal elements of this reduced density matrix are directly given by the oscillation probabilities $P_{\mu\mu}$, $P_{\mu e}$, and $P_{\mu\tau}$, which satisfy probability conservation.

For a bipartite system with effective dimension $n=2$, the predictability $\mathcal{P}^2$ is defined in terms of the diagonal elements of the reduced density matrix, as given in Eq.~(\ref{eq:def-predictability}). Applying this definition to $\rho^{\mu}_{e\tau}$ in Eq.~(\ref{eq:reducedDM}), we obtain
\begin{equation}
\mathcal{P}^2
=2\sum_{i=1}^{4}\rho^{\mu}_{{e\tau}_{ii}}-1
= 2\left(P_{\mu\mu}^{2}+P_{\mu e}^{2}+P_{\mu\tau}^{2}\right)-1\,.
\end{equation}
This expression demonstrates that predictability is entirely determined by the measurable survival and transition probabilities, providing a clear physical interpretation: $\mathcal{P}^2$ quantifies the degree to which the neutrino retains flavor information as it propagates, in direct correspondence with the underlying oscillation dynamics.
%%%%%%%%%
\section{Calculation of I-concurrence for three-flavor neutrino oscillation}
\label{appendV}
Visibility is closely related to the coherence, which is given by the Hilbert-Schmidt coherence term, $|\rho_{ij}|^2$, associated with the off-diagonal elements of the reduced density matrix. In the case of the three-flavor neutrino, it is defined by the nonlocal coherence shared by the subsystem $\mu$ (detector state) with $\{\nu_e,\nu_\tau\}$, which is equal to the sum of the bipartite coherence that $\nu_\mu$ shares with $\nu_e$ and $\nu_\tau$ separately. Therefore, $\mathcal{V}^2=\mathcal{V}^{2}_{\mu|e\tau}=\mathcal{V}^2_{\mu e}+\mathcal{V}^2_{\mu\tau}$. Starting from the off-diagonal elements of the reduced density matrix in Eq.~(\ref{eq:reducedDM}), and restricting to a bipartite partition ($n=2$), the expression for the visibility in the case of three-flavor neutrino oscillations simplifies considerably. In this limit, the visibility can be written purely in terms of oscillation probabilities
as $\mathcal{V}^2
= 2\left(P_{\mu\mu}P_{\mu e} + P_{\mu\mu}P_{\mu\tau}\right).$
\section{Calculation of I-Concurrence $(\mathcal{E}^2)$ in Three-Flavor Neutrino Oscillations}
\label{appendE}
To quantify the entanglement between the detected flavor $\nu_\mu$ and the remaining flavor modes, we consider the bipartite concurrences associated with the partitions $e|\mu\tau$ and $\tau|e\mu$. Using Eq.~(\ref{reduceDMsingle}), the bipartite concurrences can be
expressed directly in terms of neutrino oscillation probabilities. In
particular, the concurrence associated with the partition
$e\,|\,\mu\tau$ depends on the transition probability $P_{\mu e}$ 
while that for the partition $\tau\,|\,\mu e$ is governed by
$P_{\mu\tau}$. Explicitly, one finds 
\begin{equation}
C^2_{e|\mu\tau} = 4P_{\mu e}\left(1-P_{\mu e}\right),
\quad 
C^2_{\tau|\mu e} = 4P_{\mu\tau}\left(1-P_{\mu\tau}\right).
\end{equation}
Since our focus is on the correlations between the propagation states $\nu_e$ and $\nu_\tau$ with the detection state $\nu_\mu$, the total entanglement content of the system can be characterized by the sum of these two contributions. This leads to the definition of the I-concurrence as
\begin{eqnarray}
\mathcal{E}^2 
&=&\frac{1}{2}\left( C^2_{e|\mu\tau} + C^2_{\tau|\mu e} \right)  \nonumber \\
&=& 2 \left(P_{\mu\mu}P_{\mu e} + P_{\mu\mu}P_{\mu\tau}
+ 2P_{\mu e}P_{\mu\tau}\right) \,.
\end{eqnarray}
The I-concurrence thus provides a compact and experimentally accessible measure of the total bipartite entanglement shared between the detected flavor and the remaining flavor modes in the three-flavor neutrino oscillation framework.
\bibliographystyle{apsrev4-2}
\bibliography{triality}

\end{document}